# Social Requirements for Virtual Organization Breeding Environments

Jan Świerzowicz and Willy Picard,

Department of Information Technology, Poznan University of Economics,
Mansfelda 4, 60-854 Poznań, Poland
{jswierz, picard}@kti.ue.poznan.pl

**Abstract.** The creation of Virtual Breeding Environments (VBE) is a topic which has received too little attention: in most former works, the existence of the VBE is either assumed, or is considered as the result of the voluntary, participatory gathering of a set of candidate companies. In this paper, the creation of a VBE by a third authority is considered: chambers of commerce, as organizations whose goal is to promote and facilitate business interests and activity in the community, could be good candidates for exogenous VBE creators. During VBE planning, there is a need to specify social requirements for the VBE. In this paper, SNA metrics are proposed as a way for a VBE planner to express social requirements for a VBE to be created. Additionally, a set of social requirements for VO planners, VO brokers, and VBE members are proposed.

## 1 Introduction

A large variety of Collaborative Networks (CN) has emerged lately as a result of the challenges faced by both the business and scientific worlds [1]. Sanchez [2] defines Virtual Organization (VO) as a set of independent organizations that share resources and skills to achieve its mission or goal. The concept of VO Breeding Environment (VBE) has been proposed by the ECOLEAD project as a way to foster the creation of VOs [3]. A VBE is a pool of institutions that have both the potential and the will to cooperate with each other through the establishment of a "base" long-term cooperation agreement and interoperable infrastructure. When a business opportunity is identified by one member (acting as a broker), a subset of these organization can be selected and thus forming a VO [2].

In most former works, the existence of the Virtual Breeding Environment is either assumed, or is considered as the result of the voluntary, participatory gathering of a set of candidate companies. In this paper, the creation of a VBE by a third authority is considered. Organizations such as Chambers of Commerce (CoC) seem to be good candidates as institutions that may be involved in VBE creation process. CoCs usually bring together companies working in the same industry (often in the same geographical area). According to World Chambers Network [4], there are over 14,000 registered Chambers of Commerce and Industry (CCI) which in turn represent over 40 million member businesses worldwide. CoC, as organizations whose goal is to promote and facilitate business interests and activity in the community, could be good candidates for exogenous VBE creators.

Creation of VBEs, similarly to creation of VOs, requires strategic and management decision-making processes substantially different from those in traditional organizations [5]. Various aspects have to be addressed during VBE planning, from technological, organizational, economic, to legislative, psychological, and cultural ones [6]. Having in mind these aspects, the three components of CNs identified by Bifulco and Sanotoro [7] for the case of PVCs should be addressed by the planner: a VBE planner should determine a set of requirements based on business (e.g. income of potential member), knowledge (e.g. ERP used by potential member) and social aspects (e.g. number of organizations that potential member collaborates with). It should be notice that modeling these requirements requires both models and methodologies to define the needs and goals of the VBE planner. To our best knowledge, no model for social requirements for a VBE to be created currently exists.

In this paper, SNA metrics are proposed as a way for a VBE planner to express social requirements for a VBE to be created. The paper is organized as follows. In section 2, the concept of social requirements is briefly introduced, along with common SNA metrics and a short example. In section 3, an approach to social requirements used for VBE planning is presented. Section 4 concludes the paper.

## 2 Social Requirements

### 2.1 Social Network Analysis

A social network is a graph of nodes (sometimes referred as actors), which may be connected by relations (sometimes referred as ties, links, or edges). Social Network Analysis (SNA) is the study of these relations [8].

An important aspect of SNA is the fact that it focused on the how the structure of relationships affects actors, instead of treating actors as the discrete units of analysis. SNA is backed by social sciences and strong mathematical theories like graph theory and matrix algebra [9], which makes it applicable to analytical approaches and empirical methods. SNA uses various concepts to evaluate different network properties.

Recently, numerous networking tools have been made available to individuals and organizations mainly to help establishing and maintaining virtual communities. The common characteristic to all of them is that members build and maintain their own social networks, which are, then, connected to other networks through hubs (individuals that are members of two ore more networks) [5].

## 2.2 SNA Common Metrics

There are several types of measures for assessment of properties for a particular node, a group of nodes, or the whole network [10]. The most common metrics for SNA are [11–13]:

- **Size** – the size of the network is the number of nodes in a given structure,
- **Average path length** – the average of distances between all pairs of nodes,
- **Density** – the proportion of ties in a network relative to the total number possible relations,
- **Degree** – the number of ties of an actor,
- **Closeness** − the inverse of the sum of the shortest distances between each individual and every other person in the network,
- **Eccentricity** – the maximum of the shortest paths to other nodes in the network; indicates how far given node is from the furthest one in the network,
- **Neighborhood size** – the number of other actors to whom a given actor is adjacent, i.e. has a direct path,
- **Reciprocated ties density** – the ratio of the number of ties that are bidirectional to the neighborhood size.

## 2.3 Social Requirements as Reversed SNA

Social Network Analysis may used to examine a given network by evaluating some of its properties. Social requirements may be considered as the reverse approach: social requirements may be used to define some properties of a network and their associated expected values, that may then be used to check if an existing network satisfies these social requirements. It should be noticed that social requirements are usually at a higher level of abstraction than SNA metrics, and therefore, a "translation" phase between social requirements and SNA metrics is usually required.

To illustrate the concept of social requirements, let assume that a wholesaler entering the market is planning the structure of his social network. In table 1, the social requirements she/he defined during her/his network planning are presented, together with associated SNA metrics and expected values.

**Table 1.** Social Requirements of wholesaler (example)

| Social requirement | SNA metrics | Expected value |
|---|---|---|
| I want three distributors | Size of the network | =4 (including main actor) |
| Distributors must be my direct friends | Shortest path between main actor and a member | =1 |
| Distributors must not know each other directly | Shortest path between any member | >1 |
| Distributors must have at least one business partner except me | Neighborhood size | >1 |

Social business connections of the wholesaler are presented in Fig. 1, with 9 different actors connected to the actor A representing the wholesaler. The social requirements presented in table 1 define the structure of networks that would socially satisfy the wholesaler. The network consisting of actors A, F, J, I does not satisfy the wholesaler, as actor J does not meet the last requirement (his neighborhood size equals 1). On the contrary, the network consisting of actors A, F, C, and E is acceptable, as all social requirements are met in this case.

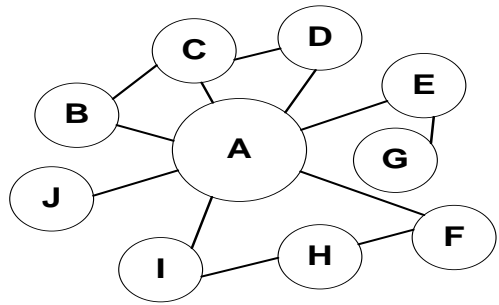

**Fig. 1.** An example of social network: the wholesaler is the A node.

It should be note that the set of networks satisfying a set of social requirements may be empty (if the social requirements are too strict), may contain one network or many networks (if the social requirements are too vague).

## 3 Social Requirements for VBEs

### 3.1 Generic Social Requirements for VBEs

Social requirements described in the former section may be an important part of VBE planning, especially during the planning of social aspects of the VBE. One should notice that, while each VBE requires an individual approach, there is a set of social requirements that are common to all VBEs, such as:

- **Size** –every VBE planner must specify at least minimal size, with 3 being the common minimal size of all VBEs. In some cases, it may be worth defining a maximal size for the planned VBE;
- **Density** – one of the main assumptions of the VBE is that the partner are interconnected (density must be at least at the level of 50%);
- **Eccentricity** – cannot be too high, whilst agile VO forming requires fast and least (if at all) mediated communication.

## 3.2 VBE Roles and Social Requirements

Social requirements may not only be defined in a generic manner as presented in the former subsection, but may also encompass the characteristics of various VBE roles formerly identified by the ECOLEAD project [3]:

- **VBE Member:** the basic role played by those organizations that are registered at the VBE and are ready to participate in the VBE activities. As regards social requirements, a VBE member cannot be a passive/isolated actor in a network, i.e. a VBE member should be at least either a sender or a receiver of information. Such a social requirement may be "translated" in terms of SNA metrics as a constraint on its density

*The inbound density or outbound density of a VBE member should be higher than 50%.*

- **VO Planer**: a role performed by a VBE actor that in face of a new collaboration opportunity, identifies the necessary competencies and capacities, selects an appropriate set of partners, and structures the new VO. As regards social requirements, a VO planner should have a good knowledge of the members of the VBE, i.e. a VO planner should have a higher level of connectivity than average VBE member. Such social requirements may be translated in terms of SNA metrics as constraints on its inbound and outbound degrees and reciprocity density.

*Inbound degrees, outbound degrees and the reciprocity density of a VO planner should be higher than the average of other VBE members.*

- **VO Broker**: a role performed by a VBE actor that identifies and acquires new collaboration opportunities. As regards social requirements, a VO broker collects information. Such social requirements may be translated in terms of SNA metrics as constraints on its inbound and outbound degrees.

*Inbound degrees, outbound degrees of a VO broker should be higher than the average of other VBE members.*

## 3.3 Example of Social Requirements for VBE

To illustrate the formerly presented approach, let imagine a Chamber of Commerce that gathers 10 steel manufacturers. CoC wonders whether it makes sense to create a VBE among these manufacturers, and if so what companies should participate. Following on that, a VBE planner from the CoC defines the following social requirements for the VBE to be potentially created:

**Table 2.** Social requirements for steel manufacturers' VBE

| Requirement | Measure | Value |
| --- | --- | --- |
| The VBE should have at least 5 members | Size | ≥ 5 |
| Members must be interconnected | Density of the network | > 50% |
| At least half of the members must have a collaboration history | Reciprocated ties | > 50% |
| There must be at least one VO broker | Inbound density | > 80% |
| There must be at least one VO planner | Inbound density and Outbound density; | > 70% |
| | Reciprocated density | > 80% |

The relations among manufacturers are modeled as a network, presented in Fig. 2, which is based on [11].

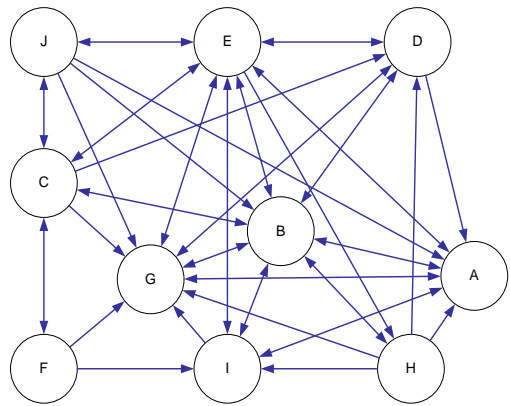

**Fig. 2.** Steel manufacturers' social network

**Table 3.** Steel manufacturers' social network matrix

|   | A | B | C | D | E | F | G | H | I | J |
|---|---|---|---|---|---|---|---|---|---|---|
| **A** | X | 1 | 0 | 0 | 1 | 0 | 1 | 0 | 1 | 0 |
| **B** | 1 | X | 1 | 1 | 1 | 0 | 1 | 1 | 1 | 0 |
| **C** | 0 | 1 | X | 1 | 1 | 1 | 1 | 0 | 0 | 1 |
| **D** | 1 | 1 | 0 | X | 1 | 0 | 1 | 0 | 0 | 0 |
| **E** | 1 | 1 | 1 | 1 | X | 0 | 1 | 1 | 1 | 1 |
| **F** | 0 | 0 | 1 | 1 | 1 | X | 1 | 0 | 1 | 0 |
| **G** | 0 | 1 | 0 | 1 | 1 | 0 | X | 0 | 0 | 0 |
| **H** | 1 | 1 | 0 | 1 | 1 | 0 | 1 | X | 1 | 0 |
| **I** | 0 | 1 | 0 | 0 | 1 | 0 | 1 | 0 | X | 0 |
| **J** | 1 | 1 | 1 | 0 | 1 | 0 | 1 | 0 | 0 | X |

The graph presented in Fig. 2 may be represented by the matrix in Table 3, where columns correspond to inbound ties (sender of information), and rows correspond to outbound ties (receiver of the information), note that self-ties are ignored. To simplify computations only binary measures are taken into consideration, i.e. the intensity of information flow is not taken into account, just the fact that a relation exists (represented in matrix by "1") or not (represented in matrix by "0").

Let check if the whole network satisfies the social requirements defined for the steel manufacturers' VBE.

The first requirement concerning the size is obviously satisfied as the required size is 5 while there are 10 manufacturers.

The second requirement concerns the density of the network which is expected to be more than 50%. Since there are 10 actors in a network, there are 90 possible connections, i.e. $n \times (n-1)$, where n is the size of the network. The actual number of ties is 51 which means that the density of the network equals 56%. The second requirement is therefore satisfied.

The third requirement concerns the collaboration history of VBE potential members, with a number of reciprocated ties expected to be greater than 50% of the number of ties of the whole network. With 51 being the total number of ties and 19 reciprocated ties, the number of reciprocated ties is $(19 \times 2) / 51 = 75\%$. The third requirement is therefore satisfied.

The forth and fifth requirements concerns the existence of at least one VO broker and one VO planer in the VBE. These requirements are related with inbound and outbound densities, as well as to reciprocated ties for a given manufacturer. Outbound density is a measure of the contribution to the network (i.e. an actor sends information to most actors), while inbound density is a measure of use of the network by an actor (i.e. an actor receives information from other actors). Table 4 presents values for each actor.

**Table 4.** Outbound, inbound and reciprocated densities for steel manufacturers

| Actor | Outbound density | Inbound density | Reciprocated density |
|-------|------------------|-----------------|----------------------|
| A | 0,44 | 0,78 | 4 (58%) |
| B | 0,78 | 0,89 | 7 (88%) |
| C | 0,67 | 0,44 | 4 (67%) |
| D | 0,44 | 0,56 | 3 (50%) |
| E | 0,89 | 0,89 | 8 (100%) |
| F | 0,33 | 0,11 | 1 (33%) |
| G | 0,33 | 1,00 | 4 (44%) |
| H | 0,67 | 0,22 | 2 (33%) |
| I | 0,33 | 0,56 | 3 (50%) |
| J | 0,56 | 0,22 | 2 (40%) |

About outbound density, actor E sends information to all but actor F, and its outbound density – 89% – is highest in the network. As a consequence, actor E has the highest potential to be influential. Actors B and E are the two only actors with an outbound density higher than 70% (cf. the fifth requirement).

About inbound density, the actors A, B, E, and G have inbound densities higher than 70% (cf. the forth requirement). All these actors but actor A have inbound densities higher than 80% (cf. the fifth requirement). As a consequence, actors B, E and G are potential candidates to the role of VO brokers. Therefore the forth requirement is satisfied.

From their outbound and inbound densities, only the manufacturer B and E are potential VO planner, under condition that their reciprocated density is greater than 80 % (cf. the fifth requirement). This condition is satisfied for actors B and E. Therefore the fifth requirement is satisfied.

As a conclusion, all five requirements for the steel manufacturers are satisfied. Additionally, the social requirements concerning VO planers and VO brokers defined in section 3.2 are more lenient than the forth and fifth requirements. Therefore, the social requirements related with VBE roles for VO planers and VO brokers defined in section 3.2 are satisfied.

But the social requirements concerning VBE members defined in section 3.2 are not satisfied by actor F. The number of ties of actor F is 4 (1 inbound and 3 outbound), while the number of potential ties of actor F is twice the number of remaining actors, i.e. 18 (9 inbound and 9 outbound). Therefore, the outbound density of actor F equals 3 / 9 = 33% and inbound density equals 1 / 9 =11%. Neither of these results exceeds 50%, therefore density requirement for VBE members defined in section 3.2. is not satisfied for actor F.

As a conclusion, the steel manufacturers' social network presented in Figure 2 satisfies the social requirements defined in table 2, but some actors do not satisfy social requirements concerning VBE roles. Therefore, the considered social network should not be casted into a VBE.

A further step could be the removal of actors that do not satisfy social requirements for VBE roles, e.g. actor A. Next, it should be checked if the resulting social network satisfies all social requirements.

## 4   Conclusions

The main contribution presented in this paper is twofold: first, the idea of exogenous VBE creation is proposed; second, the concept of social requirements for VBEs is introduced. The VBE creation process has currently been the subject of little work as it is usually assumed that either the VBE exists or that the VBE is the result of the participatory gathering of voluntary companies, eventually asking outside institutions for support [14].

Second, to our best knowledge, the use of SNA as a basis for modeling social requirements for VBEs is a novel approach to VBE modeling. Using Social Network Analysis methods to examine abovementioned requirements enables quantitative specification of characteristics of VBE often described in qualitative way.

Among future works, the concepts presented in this paper should be formally defined. Next, a methodology to translate social requirements into appropriate SNA metrics is still to be proposed. Finally, algorithms for the identification, within a given network of organizations, of sub-networks that fulfill a given set of social requirements need to be developed.

**Acknowledgments.** This work has been partially supported by the Polish Ministry of Science and Higher Education within the European Regional Development Fund, Grant No. POIG.01.03.01-00-008/08.